\documentclass[a4paper]{spie}
  
\usepackage{amsmath,amsfonts,amssymb}
\usepackage{graphicx}
\usepackage[colorlinks=true, allcolors=blue]{hyperref}
\usepackage{wasysym}
\usepackage{doi}

\label{title}
\title{Development of TIFUUN: Terahertz Integral Field Units with Universal Nanotechnology}

\author[a]{Akira Endo}
\author[b]{Tom J. L. C. Bakx}
\author[a,c,d]{Jochem J. A. Baselmans}
\author[e]{Dries Boleij}
\author[a]{Stefanie A. Brackenhoff}
\author[f,g]{Bernhard R. Brandl}
\author[h]{Martino Calvo}
\author[a,c]{Shahab O. Dabironezare}
\author[e]{Hans van der Does}
\author[i]{Rei Enokiya}
\author[j]{Sho Fujisawa}
\author[k]{Shinji Fujita}
\author[l]{Enrico Garaldi}
\author[e]{Wouter Gregoor}
\author[j]{Masato Hagimoto}
\author[a]{Davit Hakobyan}
\author[m]{Angelina Harke-Hosemann}
\author[c]{Robert Huiting}
\author[k,i,l]{Shiro Ikeda}
\author[n,o]{Reinier M. J. Janssen}
\author[c]{Kenichi Karatsu}
\author[c]{Nick de Keijzer}
\author[m,p]{Kotaro Kohno}
\author[j]{Takumi Kojima}
\author[a]{Alkistis Kyriakidou}
\author[a]{Louis H. Marting}
\author[l]{Tomotake Matsumura}
\author[e]{Cory Meijneke}
\author[i,q]{Tetsuhiro Minamidani}
\author[a]{Arend Moerman}
\author[p,r,t]{Kana Moriwaki}
\author[h]{Alessandro Monfardini}
\author[s,i]{Kanako Narita}
\author[u]{Yuri Nishimura}
\author[p]{Erika Ogata}
\author[a]{Leon G. G. Olde Scholtenhuis}
\author[a]{Tristan Oude Essink}
\author[a]{Jim R. Piek}
\author[f]{Matus Rybak}
\author[v]{Kana Sakaguri}
\author[i]{Seiichi Sakamoto}
\author[c]{Aurora Simionescu}
\author[a,w]{Nikita A. Soshnin}
\author[x]{Tatsuya Takekoshi}
\author[j]{Yoichi Tamura}
\author[x]{Akio Taniguchi}
\author[c]{David J. Thoen}
\author[a]{Sten Vollebregt}
\author[c]{Lingyu Wang}
\author[f]{Paul P. van der Werf}
\author[c]{Stephen J. C. Yates}
\author[l,r,t]{Naoki Yoshida}
\author[a]{Silvia Zhang}

\affil[a]{Faculty of Electrical Engineering, Mathematics and Computer Science, Delft University of Technology, Mekelweg 4, 2628 CD Delft, the Netherlands.}
\affil[b]{Department of Physics and Astronomy, Chalmers University of Technology, Chalmersplatsen 4 412 96 Gothenburg, Sweden}
\affil[c]{SRON Space Research Organisation Netherlands, the Netherlands}
\affil[d]{University of Cologne, Cologne, Germany}
\affil[f]{Leiden Observatory, Leiden University, Einsteinweg 55, 2333 CC Leiden, the Netherlands}
\affil[g]{Faculty of Aerospace Engineering, Delft University of Technology, Kluyverweg 1, 2629 HS Delft, the Netherlands.}
\affil[e]{Electronic and Mechanical Support Division (DEMO), Delft University of Technology, Mekelweg 4, 2628 CD Delft, the Netherlands.}
\affil[h]{Institut N\'eel, CNRS and Universit\'e Grenoble Alpes (UGA), France}
\affil[i]{National Astronomical Observatory of Japan (NAOJ), National Institutes of Natural Sciences (NINS), 2-21-1, Osawa, Mitaka, Tokyo 181-8588, JAPAN}
\affil[j]{Department of Physics, Graduate School of Science, Nagoya University, Furo, Chikusa, Nagoya, Aichi 464-8602, Japan}
\affil[k]{The Institute of Statistical Mathematics, 10-3 Midori-cho, Tachikawa, Tokyo 190-8562, Japan}
\affil[l]{Kavli Institute for the Physics and Mathematics of the Universe (Kavli IPMU, WPI), the University of Tokyo, 5-1-5 Kashiwanoha, Kashiwa, Chiba 277-8583, Japan}
\affil[m]{Institute of Astronomy, Graduate School of Science, The University of Tokyo, 2-21-1 Osawa, Mitaka, Tokyo 181-0015, Japan}
\affil[n]{Jet Propulsion Laboratory, California Institute of Technology, 4800 Oak Grove Drive, Pasadena, CA 91109, USA}
\affil[o]{Department of Astronomy, California Institute of Technology, 1200 E California Blvd, Pasadena, CA 91125, USA}
\affil[p]{Research Center for the Early Universe, Graduate School of Science, The University of Tokyo, 7-3-1 Hongo, Bunkyo, Tokyo 113-0033, Japan}
\affil[q]{Graduate Institute for Advanced Studies, SOKENDAI 2-21-1, Osawa, Mitaka, Tokyo 181-8588, JAPAN}
\affil[r]{Department of Physics, Graduate School of Science, The University of Tokyo, 7-3-1 Hongo, Bunkyo-ku, Tokyo 113-0033, Japan}
\affil[s]{Department of Astronomy, Graduate School of Science, The University of Tokyo, 7-3-1 Hongo, Bunkyo-ku, Tokyo 113-0033, Japan}
\affil[t]{RIKEN Center for Advanced Intelligence Project, 1-4-1 Nihonbashi, Chuo, Tokyo 103-0027, Japan}
\affil[u]{Faculty of Pure and Applied Sciences, University of Tsukuba, 1-1-1 Tennodai, Tsukuba, Ibaraki 305-8577, Japan}
\affil[v]{Department of Physics, Yale University, 217 Prospect St, New Haven, CT 06520, USA}
\affil[w]{STRATAL SYSTEMS, Amstelveen, the Netherlands}
\affil[x]{Kitami Institute of Technology, 165 Koen‑cho, Kitami, Hokkaido 090‑8507, Japan}

\authorinfo{Further author information:
	A. Endo: E-mail: a.endo@tudelft.nl}

\begin{document} 
\maketitle

% Abstract
\begin{abstract}
\label{abstract}
TIFUUN (THz Integral Field Units with Universal Nanotechnology) is an ultra-wideband mm-submm wave imaging spectrometer that capitalizes on the highly scalable integrated superconducting spectrometer technology. TIFUUN has two slots for integral field units (IFUs), which can jointly be optimized as open-hardware for each astronomical observation in terms of spatial and spectral coverage. These IFUs can have observation frequencies in the range of 90--360 GHz, with spectral resolution up to $R\equiv F/\Delta F \le 1,000$, with up to $\sim$18,000 kinetic inductance detectors (shared by the two IFUs with a flexible ratio). The ultra-wide 4:1 (2 octave) bandwidth optics fits in a remarkably compact volume, by means of thin silicon lenses and a high chief ray angle design. The first pair of IFUs are being developed for the SUBLIME (Study of the Universe By Line Intensity Mapping Experiments) experiment that aims to map $[\mathrm{CII}]$ emission at redshift $\sim$6 to trace the cosmic large-scale structure and the buildup of galaxies during reionization, using TIFUUN on the ASTE 10-m telescope. The scalability, flexibility and compactness makes TIFUUN a highly compatible and portable system suited also for upcoming telescope facilities in the vicinity, such as FYST and AtLAST/LST.
\end{abstract}


\section{INTRODUCTION}
\label{sec:intro}

% Figure Open HW 
\begin{figure} [ht]
\begin{center}
\includegraphics[width=170mm]{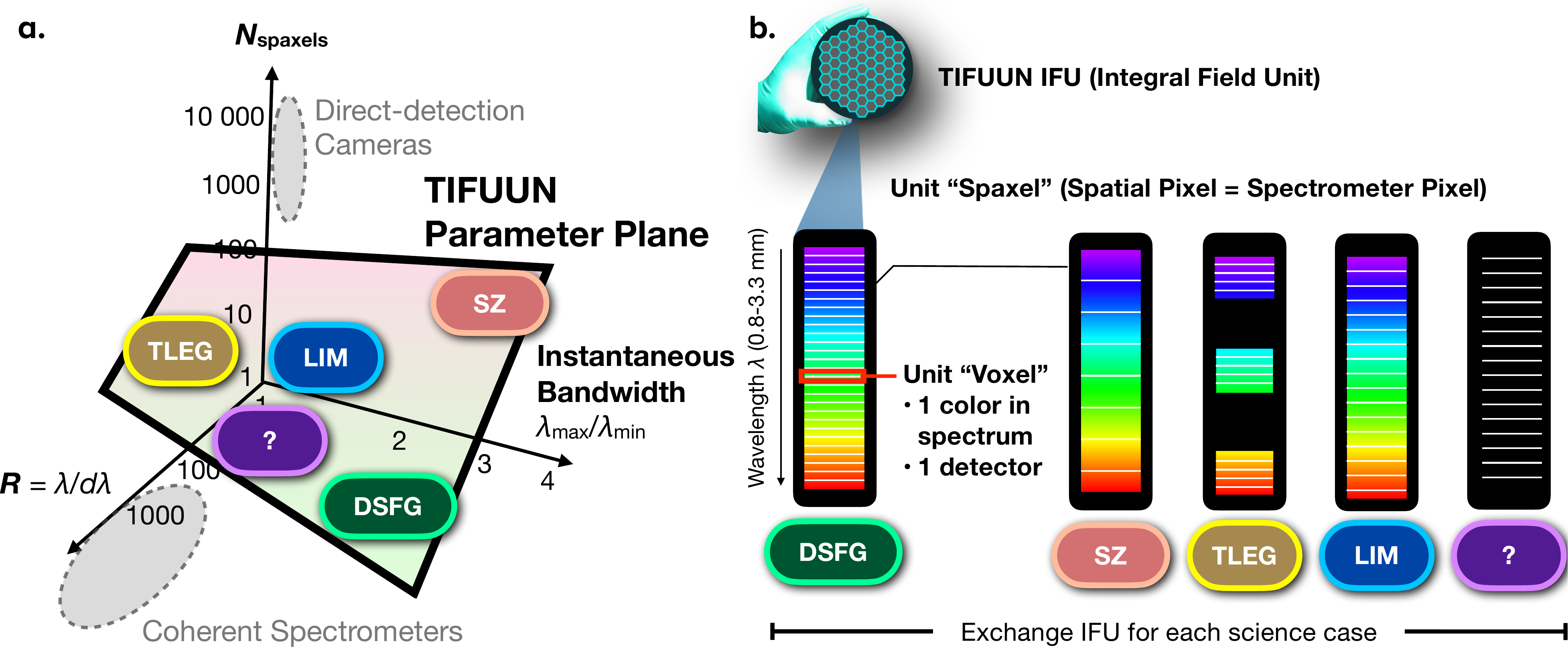}
\end{center}
\caption[Open Hardware]
{\label{fig:open-hw} 
\textbf{a.} TIFUUN will open a vast discovery space with a variety of observing capabilities in the {resolution, bandwidth, field-of-view} plane, which is hard to access by conventional technology (coherent receivers and direct-detection cameras, indicated in grey shades.) Science cases: DSFG = Dusty Star-Forming Galaxies; SZ = Galaxy cluster physics with the Sunyaev-Zeldovich effect; TLEG = Terahertz Line Emitting Galaxy survey; LIM = Large-scale structure of the cold interstellar matter probed by Line Intensity Mapping. \textbf{b.} The TIFUUN Integral Field Unit (IFU) is a superconducting circuit patterned on a silicon wafer. On the IFU there is a 2D array of ``spaxels,'' each of which is an independent spectrometer measuring a point on the sky. The IFU design is optimized for each science case indicated by the common labels in panels a and b.}
\end{figure} 

Spectral imaging with a wide bandwidth ($>$octave) and moderate spectral resolution ($R\equiv F/\Delta F \sim 10^2$--$10^3$) in the millimeter-submillimeter (mm-submm) band is key to addressing a broad range of astrophysical questions, including the growth of large-scale structure\cite{Snowmass,LIM2026}, the dynamics of galaxy clusters\cite{Mroczkowski2019}, and the cosmic history of star formation\cite{KohnoSPIESpectrograph2020}.  
However, current observational technologies are limited by a fundamental dichotomy between coherent receivers and direct-detection cameras, as shown in the dashed areas of Fig. \ref{fig:open-hw}a.
On one side, coherent spectrometers have reached very high sensitivity with the ALMA and NOEMA interferometers, delivering images and spectra with exquisite resolution, but for a minute cosmological volume because of the limited bandwidth and field of view. 
On the other side, direct-detection cameras have revealed the ubiquitous importance and abundance of dust-obscured processes in the early Universe, motivating the entire field of mm-submm astronomy to a large degree\cite{Casey2014}, but fail to elucidate the physical processes within and among those galaxies because of the lack of spectroscopic diagnosis, especially redshift information. 
Quasioptical direct detection spectrometers have tried to fill this gap\cite{Stacey2011,CONCERTO2020}, but they are limited in scalability because the size of the (cryogenic) optics scales linearly with the wavelength (and spectral resolution, for non-resonant spectrometers).

The integrated superconducting spectrometer (ISS)\cite{Endo2019NatureAstron,Karatsu2026DESHIMA2} is a highly scalable direct-detection imaging spectrometer design that can open the vast, uncharted discovery space between the two existing technologies as shown in Fig. \ref{fig:open-hw}a. By integrating the superconducting filterbank circuit and an array of kinetic inductance detectors (KIDs) on the same chip, the ISS can be made so small that it fits under a single lenslet with a diameter of $\sim$1 cm, as shown in Fig. \ref{fig:IFU}b. 
Such an ISS can form a unit cell (or ``spaxel'', short for ``spatial pixel'') to naturally form a 3D imaging spectrometer, also known as an integral filed unit (IFU).
The DESHIMA instrument\cite{Endo2019NatureAstron,Karatsu2026DESHIMA2} uses a single-spaxel ISS for astronomical observations on the ASTE 10-m telescope and showed photon-noise limited sensitivity\cite{Endo2019NatureAstron}, whereas imaging spectrometers with multiple spaxels are being developed and tested\cite{Superspec2022,SPT-SLIM2025Benson}.

Here we present the conceptual design and technology development of TIFUUN: THz Integral Field Units with Universal Nanotechnology.
TIFUUN is designed to be a powerful yet flexible instrument, such that the spatial coverage (number and arrangement of the spaxels), the spectral coverage, and the spectral resolution can be optimized within a wide range for each IFU, tailored to each astronomical observation as illustrated in Fig. \ref{fig:open-hw}b. 
Furthermore, TIFUUN has a compact design that makes it portable and compatible with many telescopes.
As its first scientific exploitation, a pair of IFUs is under development for the SUBLIME (Study of the Universe By Line Intensity Mapping Experiments) experiment, to conduct line intensity mapping (LIM) of the 1.9~THz $[\mathrm{CII}]$ line at  $z\sim 6$.
In this paper we describe the conceptual design of the general TIFUUN instrument system, as well as the IFUs specifically targeted towards the SUBLIME experiment.

\section{Open Hardware IFU's}
\label{sec:open-hw} 

The TIFUUN system aims to open a broad range of parameters in the \{$N_\mathrm{spaxels}$, $R$, $\lambda_\mathrm{max}/\lambda_\mathrm{min}$\} space as shown in Fig. \ref{fig:open-hw}a, to provide users with a large degree of freedom in designing the IFUs to enable a variety of observations.
The maximum number of KIDs ($N_\mathrm{KIDs}$) is targeted at 18,000, which requires 3,000 KIDs to be read out per single channel of the SpaceKIDs electronics\cite{Baselmans2017kilopixel} within the 2--4 GHz band.
These KIDs will be shared between the two IFUs: taking SUBLIME as an example, 15,000 KIDs will be allocated to the high-frequency band (H-band), whereas 3,000 KIDs will be allocated to the low-frequency band (L-band), as shown in Table \ref{tab:specs}. 
For the frequency range, the optics is for 90--360~GHz, and each IFU will typically have a bandwidth of up to 2:1 (1 octave).
Regarding the spectral resolution $R$, 500 has been demonstrated by DESHIMA and up to $\sim$1,000 could become possible, considering that the dielectric loss of amorphous silicon carbide (a-SiC) corresponds to an internal resonator quality factor $Q_\mathrm{i}$ of $\sim$10,000 at $\sim$ 200 GHz\cite{Buijtendorp2022}, though such a high $Q_\mathrm{i}$ has yet to be demonstrated in an actual bandpass filter configuration.
The number of spatial pixels is limited by the focal plane size, where the $\diameter$100-mm circle corresponds roughly to a $\diameter$7.5 arcmin diameter field-of-view on the ASTE 10-m telescope.
These figures of merits are related as $N_\mathrm{KID} \sim N_\mathrm{spaxels} \cdot R\ln (F_\mathrm{max}/F_\mathrm{min})$, showing the trade-off between the parameters.

Within these limits, the user can design IFUs that sample spatially and spectrally in their desired way.
For example, IFUs to observe the broadband Sunyaev–Zeldovich effect could prioritize on $N_\mathrm{spaxels}$ and bandwidth over spectral resolution, whereas IFUs to observe emission lines from THz line-emitting galaxies (TLEGs) would require a spectral resolution of at least a few hundred, at the cost of either $N_\mathrm{spaxels}$ or the bandwidth.
IFUs for LIM could have an optimum in between these two extreme cases.
To support potential users to propose new IFU configurations in combination with a matching science case, the \texttt{RAIMAD}\cite{RAIMAD} python package for designing TIFUUN IFUs is being developed open-source, introducing the open-hardware paradigm that is flourishing in quantum computing\cite{OpenHardware2024} to the field of superconducting detectors for astronomy.

\section{The SUBLIME experiment}
\label{sec:sublime} 

% Table Specs
\begin{table}[th]
\caption{TIFUUN and SUBLIME specs}
\label{tab:specs}
\begin{tabular}{llllll}
 & Frequency Range & $N_\mathrm{spaxels}$ & $R\equiv F/\Delta F$ & $N_\mathrm{channels}$ & $N_\mathrm{KIDs}$ \\ \hline
TIFUUN system & 90--360 GHz  &    & $\le 1,000$ &          & $\le 18,000$      \\ 
SUBLIME (H-band)    & 195--319 GHz & 61 & 500 (100)  & 246 (49) & 15,000 (3,000) \\ 
SUBLIME (L-band)    & 130--178 GHz & 19 & 500        & 158      & 3,000          \\ 
\end{tabular}
\end{table}

% Figure Sensitivity
\begin{figure} [ht]
\begin{center}
\includegraphics[width=120mm]{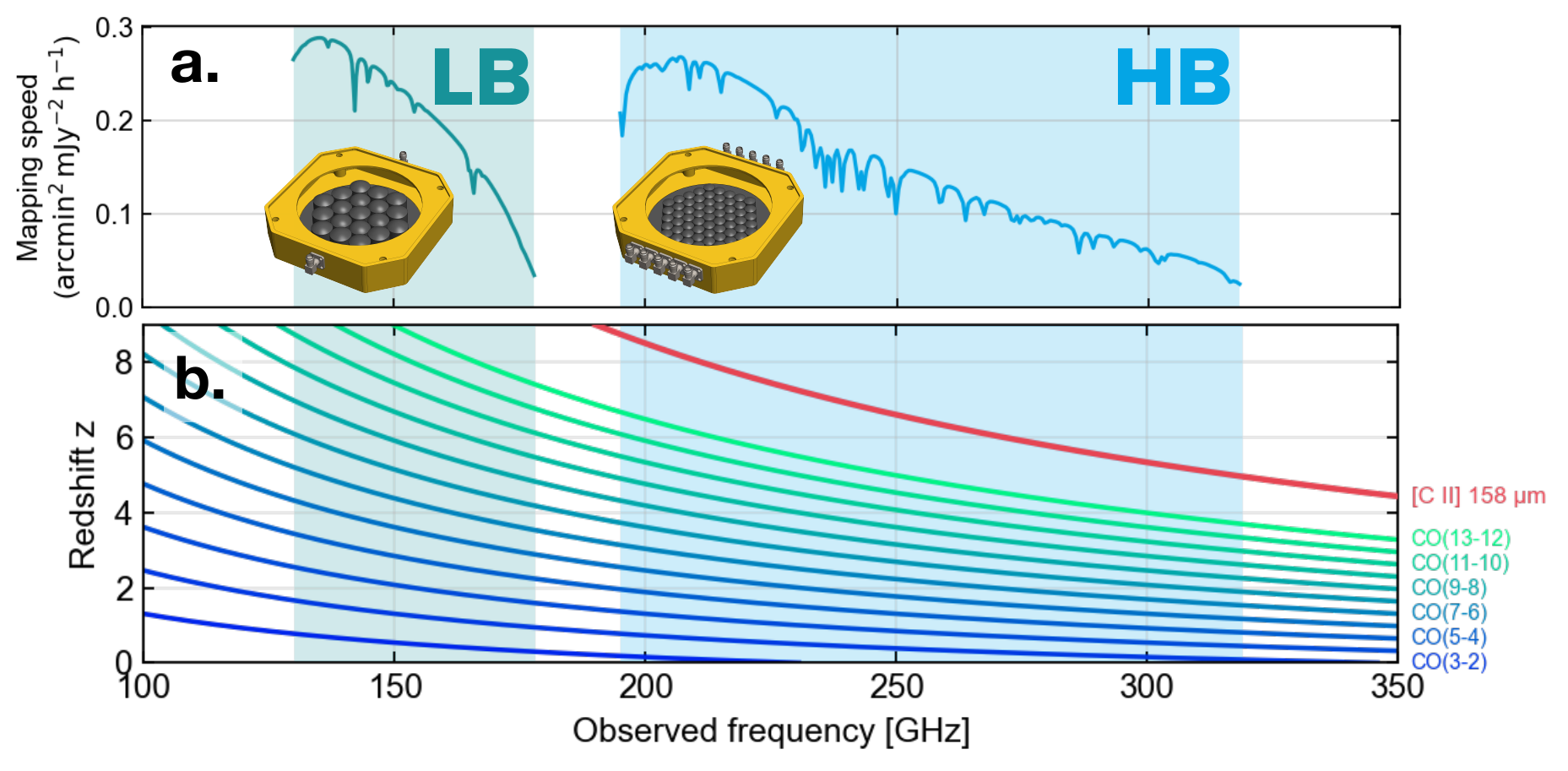}
\end{center}
\caption[Sensitivity] 
{\label{fig:sensitivity} 
\textbf{a.} Estimated mapping speed of the low-frequency band (LB) and high-frequency band (HB) IFUs on the ASTE 10-m telescope, assuming a precipitable water vapor (PWV) of 1.0 mm. Note the high frequency resolution of $R \equiv F/\Delta F =500$: the speed will improve if multiple frequency channels are averaged.  \textbf{b.} The HB will perform LIM with the $\mathrm{[CII]}$ line at $z = 4.9$--8.7, while using both bands to detect at least three CO lines from interlopers at $z>1$.}
\end{figure} 

The SUBLIME experiment aims to conduct a unified mm-submm survey that treats LIM and blind detection of TLEGs as two complementary analyses of the same 3D spectral data cube.
LIM measures spatial fluctuations of integrated line emission to trace large-scale structure statistically\cite{Snowmass}, while a blind line survey identifies discrete galaxies through their emission lines\cite{KohnoSPIESpectrograph2020,Kovacs2025ConceptCamera}. 
From the same observation, the former constrains clustering and the aggregate line luminosity density, and the latter builds an unbiased redshift catalog of luminous emitters. 
Analyzing ``both sides of the coin'' together cross-validates calibration and systematics, and turns a single observation into both precision clustering statistics and a redshifted emitter census, enabling tighter constraints on the luminosity function and cosmic star-formation history than either technique can deliver alone. 
%Moreover, the LIM measurements with LIM-TIFUUN can be cross-correlated with the forthcoming NASA SPHEREx all-sky spectroscopic survey, enabling powerful synergy between near-infrared and terahertz tracers of the large-scale structure.

For SUBLIME, we are developing two IFUs with parameters presented in Table \ref{tab:specs}. The H-band IFU will cover 195--319 GHz, targeting the $[\mathrm{C II}]$ 1.9 THz line at $z=4.9$--8.7, with a 61-spaxel focal-plane hosting 15,000 KIDs. The L-band IFU will span 130--178 GHz with 19 spaxels and 3,000 KIDs. Both arrays operate at a resolving power of $R=500$, a resolution already demonstrated by DESHIMA\cite{Karatsu2026DESHIMA2}, and well matched to LIM needs and the spectral line widths of bright TLEGs. 
A central challenge for $[\mathrm{CII}]$ LIM is contamination by lower-redshift CO rotational lines. Our strategy is to identify and remove CO interlopers through cross-correlation of multiple CO transitions measured by the L-band IFU. With 130--178~GHz and 195--319~GHz coverage, any galaxy at $z>1$ places three or more CO lines inside the bands at once, as shown in Fig. \ref{fig:sensitivity}b.

Thanks to the large (10 m) diameter of the ASTE telescope, the same observations naturally support a blind search for individual emitters with line widths characteristic of massive dusty star-forming galaxies. 
Existing and planned wide-area mm–submm surveys are often biased toward dust-rich systems because they first pre-select sources in continuum and then follow with spectroscopy. Blind spectral mapping with TIFUUN omits the continuum pre-selection entirely, delivering an unbiased, redshifted THz line-emitter survey. The H-band IFU conducts a $\mathrm{[CII]}$ emitter survey at $z=4.9–8.7$, while the L-band could test cross-correlation/stacked detection across multiple CO lines (e.g., as advocated by Kov{\'a}cs et al.~\cite{Kovacs2025ConceptCamera})

The IFU specifications for SUBLIME are presented in Table \ref{tab:specs}. The goal configuration will have $R=500$ in both bands, requiring a total of 18,000 KIDs (3,000 KIDs per RF readout chain). As an intermediate step, we plan to develop an $R=100$ IFU for the H-band, requiring 3,000 KIDs in both H-band and L-band, as indicated by parentheses in the table. This version requires reading out 1,000 KIDs per RF readout chain, which has been successfully demonstrated\cite{Baselmans2017kilopixel}.

\section{Superconducting Integral Field Units}
\label{sec:IFU} 

% Figure IFU
\begin{figure} [ht]
\begin{center}
\includegraphics[width=170mm]{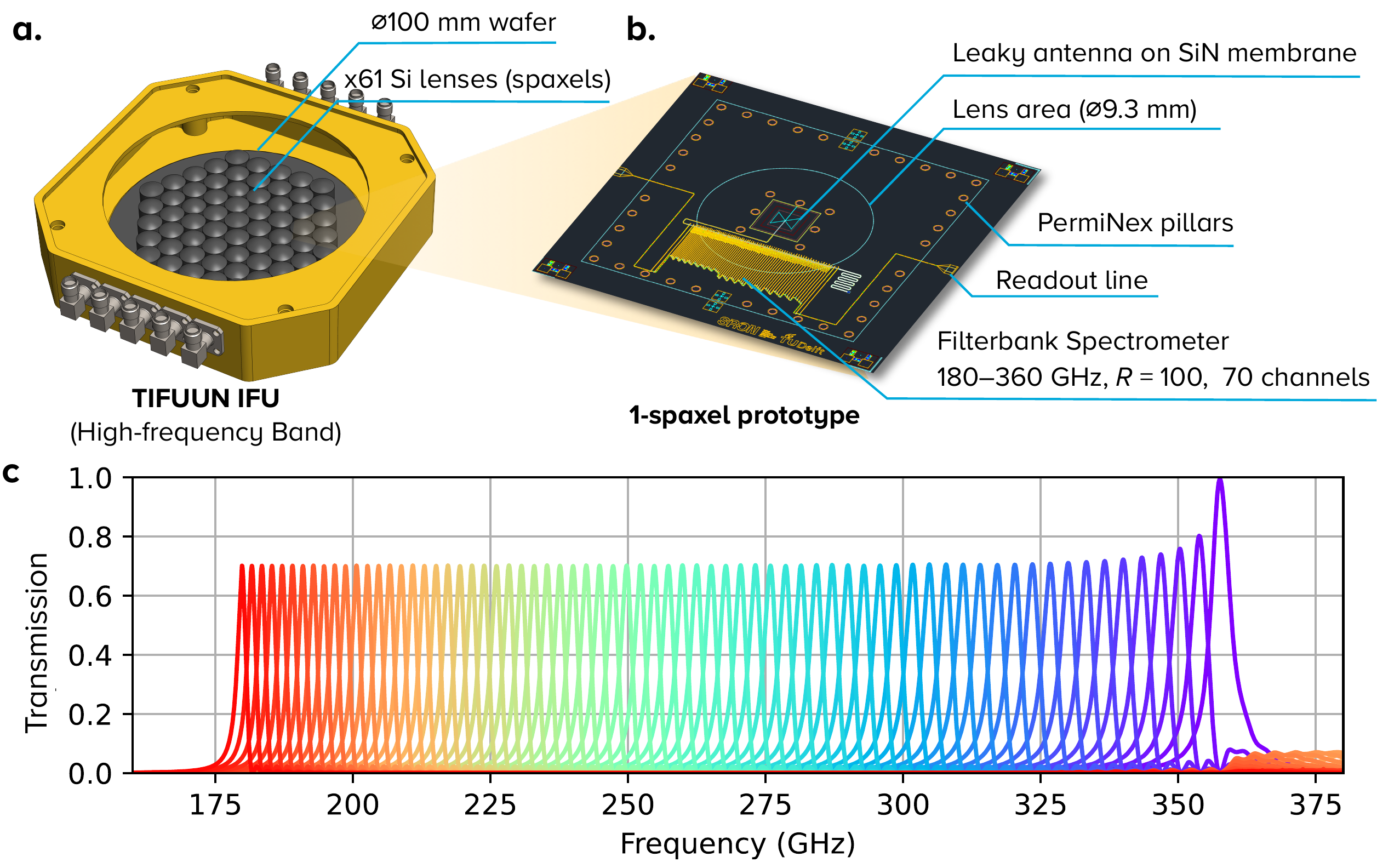}
\end{center}
\caption[IFU] 
{\label{fig:IFU} 
\textbf{a.} The Integral field unit (IFU) of TIFUUN is patterned on a $\diameter$100-mm Si wafer. Each spaxel couples to the radiation from the optics with a leaky-lens antenna. 
\textbf{b.} Under each lens of the IFU, there is an integrated superconducting spectrometer (ISS) similar to the single-spaxel prototype chip, of which the design is shown here. This is a prototype design that requires modifications to achieve the desired packing as shown in panel a. In this example that uses NbTiN/Al hybrid KIDs coplanar waveguide (CPW) technology, roughly 70 channels occupies most of the available area under a lens, motivating the development of KIDs with a smaller footprint to increase the number of channels per spaxel. The PermiNex pillars are for creating the space between the antenna and the lens\cite{Karatsu2026DESHIMA2}.
\textbf{c.} The simulated transmission (from the entrance of the filter bank to the KID) of a 70-channel, $R=100$ filterbank based on the directional filter technology demonstrated by Marting et al.\cite{Marting2026Directional}.}
\end{figure} 

The conceptual design of a TIFUUN IFU for the SUBLIME H-band is presented in Fig. \ref{fig:IFU}. 
On the $\diameter$100-mm wafer, 61 lenses are hexagonally packed to cover the $\sim$7.5-arcmin field-of-view of the ASTE telescope.
The diameter of the lenses and their spacing are optimized for maximum total mapping speed.
The lens is made of high-resistivity silicon, and requires an anti-reflection coating by either coating\cite{Karatsu2026DESHIMA2} or machining sub-wavelength structures\cite{Bueno2022LensAR}. 
At the focus of each lens, there is a leaky antenna that couples the radiation to the microstrip line that guides the signal to the filterbank.
Each channel of the filterbank is a combination of a high-efficiency directional filter\cite{Marting2026Directional} using NbTiN/SiC/NbTiN microstrip resonators, and a NbTiN/Al hybrid KID.
If conventional CPW KIDs (such as the ones used for DESHIMA\cite{Karatsu2026DESHIMA2} and AMKID\cite{AMKID_Reyes2026} ) are used, around $\sim$50 KIDs (of 2--4 GHz readout frequency) fit under the $\diameter$9.3-mm lens, considering also the space that is required for the antenna and readout line.
To integrate the 246 channels of the H-band of SUBLIME, the area per KID must be reduced by a factor of $\sim$5.
A promising option is to replace the wide NbTIN CPW with a parallel plate capacitor (PPC) that has a higher capacitance per unit area, hence reducing the required footprint of the KID\cite{Hempel-Costello2026}.
Whilst the vibrational modes in the far-infrared band determine the mm-submm loss\cite{Buijtendorp2025} (hence the maximum $R$ of a filter), the noise of a PPC KID is set by the TLS noise.
Hence, development of a deposited dielectric material (as well as the interface with NbTiN) that satisfied both requirements is key\cite{Hempel-Costello2026,Buijtendorp2022}.

\section{Optics}
\label{sec:optics} 

% Figure Optics
\begin{figure} [ht]
\begin{center}
\includegraphics[width=170mm]{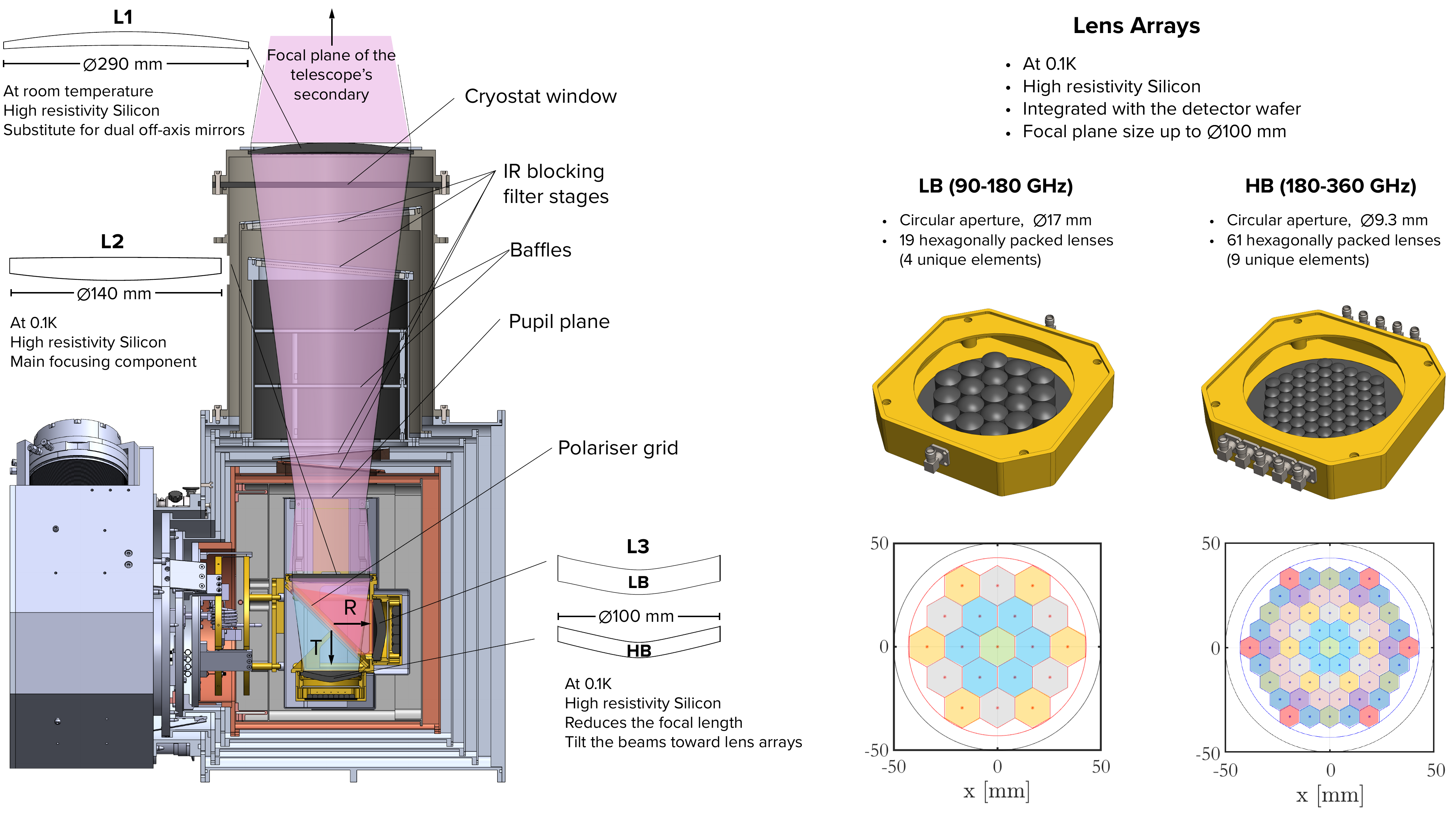}
\end{center}
\caption[Optics] 
{\label{fig:optics}
Optics design of TIFUUN. With a high chief-ray angle design, the optical chain is only $\sim$1 m tall. \textbf{a.} Lens 1 (L1) and Lens 2 (L2) are common for both bands, requiring them to perform over the full 90--360 GHz. After separating the signal in two linear polarizations, there is an additional lens 3 (L3) for each polarization to couple the beams to the IFUs. \textbf{b.} The lens/spaxel layout of each band. Each color within each array represents a unique lens shape required to comply with the high CRA optics\cite{Dabironezare2026EUCAP}. 
}
\end{figure} 

The optical configuration of TIFUUN is presented in Fig. \ref{fig:optics}.
For the given field-of-view and frequency range, the optics is remarkably compact thanks to the high chief-ray angle (CRA) design that is optimized using a sequential geometric optics technique.
The details of the optical design and the performance is presented in ref. \cite{Dabironezare2026EUCAP}.
The first lens seen from the secondary dish of the ASTE telescope, L1, is placed near the telescope focal plane in the upper receiver cabin.
L1 is at room temperature, and it also has the largest optical diameter of $\diameter$290 mm.
To reduce thermal optical loading from this lens, high resistivity silicon is a desired material, though float zone (FZ) silicon is not easily available for diameters larger than $\diameter$200 mm.
It has been reported that silicon made with the Czochralski (CZ) method can also have low millimeter-wave losses at room temperature\cite{Datta2013}, and available to up to $\diameter$450 mm.
In addition, all the lenses must have an anti-reflection (AR) structure because silicon has a large index of refraction ($n\sim3.4$).
To this end, a promising approach is machining sub-wavelength structures by laser ablation\cite{Takaku2026} or dicing \cite{Datta2013,Nitta2018}.
Inside the cryostat, there is one lens that is common for both bands (L2), and after the polarizing grid there is a final lens for each band that serve as  field lenses(L3-HB, L3-LB).

\section{Cryo-mechanical Design}
\label{sec:cryostat} 

% Figure Cryo Mechanical Cross Section
\begin{figure} [ht]
\begin{center}
\includegraphics[width=170mm]{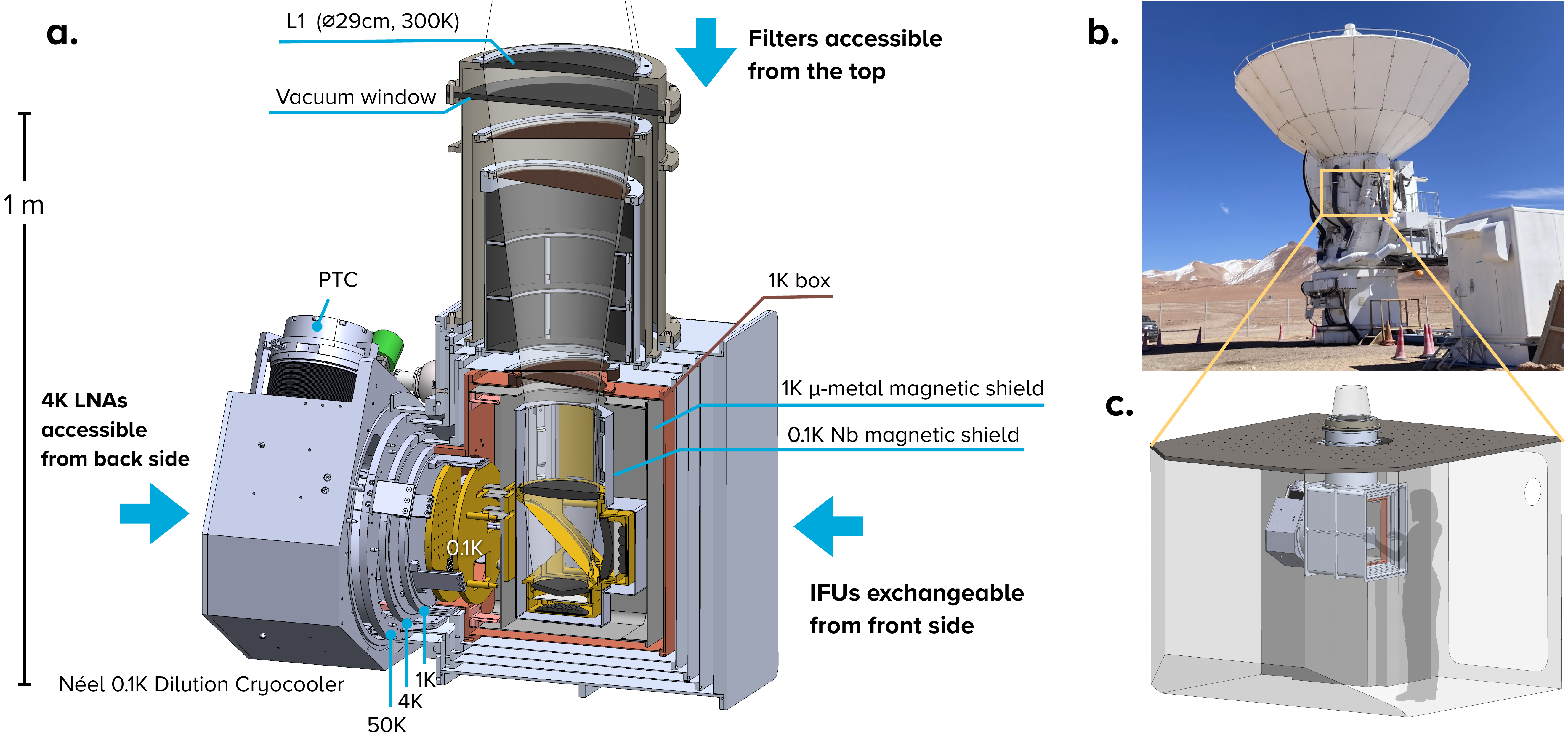}
\end{center}
\caption[Cryostat] 
{\label{fig:cryostat} 
\textbf{a.} Cross-sectional computer-aided drawing (CAD) model of the TIFUUN cryo-mechanical structure. 
\textbf{b.} ASTE 10-m telescope. The box indicates the location of the Cassegrain receiver cabin.
\textbf{c.} CAD model of the TIFUUN cryostat placed in the ASTE receiver cabin, with a person indicating how the IFUs can be exchanged by opening one side.
}
\end{figure} 

% Figure Magnetic Shielding
\begin{figure} [ht]
\begin{center}
\includegraphics[width=170mm]{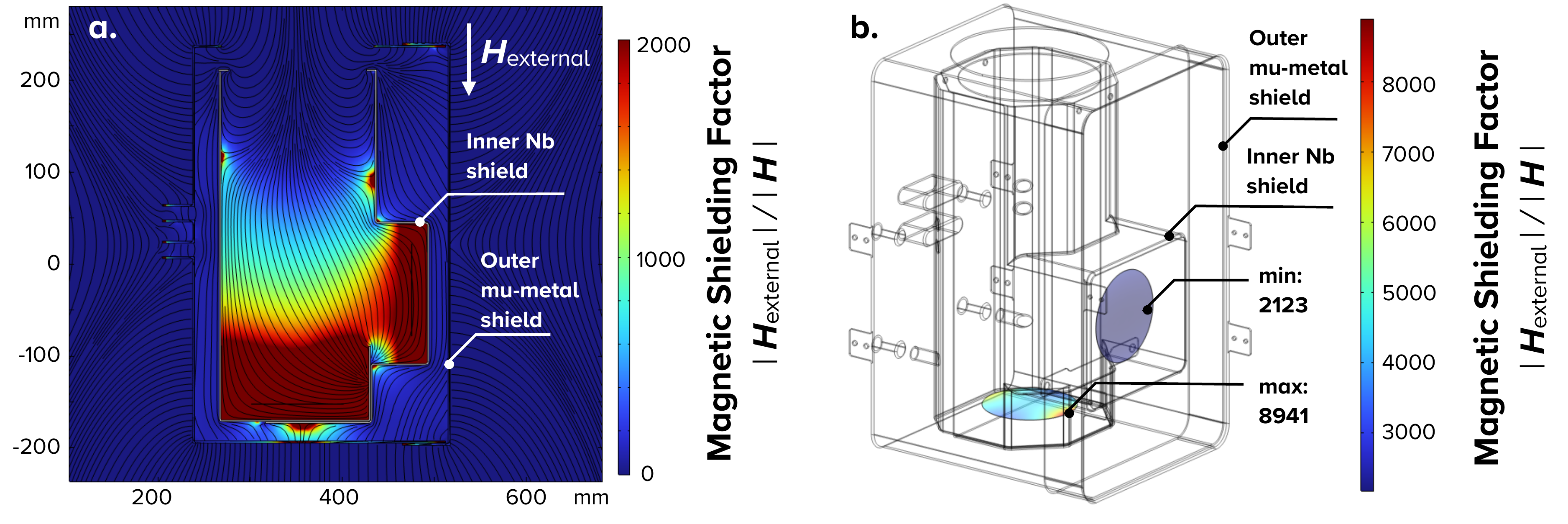}
\end{center}
\caption[Magnetic Shield]
{\label{fig:mag_shield} 
Simulation of the mu-metal/Nb double magnetic shield, using COMSOL Multiphysics. \textbf{a.} The color scale indicates the magnetic shielding factor against a magnetic field applied in the direction indicated by the arrow, which is the direction that is hardest to shield. \textbf{b.} The shielding factor over the area of the IFUs. The minimum shielding factor is $\sim$2.1$\cdot 10^3$, satisfying the requirement of $>$2.0$\cdot 10^3$ that we set.
}
\end{figure} 

The mechanical design of TIFUUN is presented in Fig. 5a.
The combination of the integrated spectrometer, the high-CRA optics, and the compact dilution refrigerator derived from CONCERTO\cite{CONCERTO2020} make the frontend system remarkably compact, with a height of $\sim$1~m.
The cryostat can be opened from 3 directions, for easy access to the IFUs, the low-noise amplifiers (LNAs), and the quasioptical filters.

The mechanical and optical design of TIFUUN began with the constraint that the cryocooler needed to be placed in a fixed side of the ASTE receiver cabin as indicated in Fig. \ref{fig:cryostat}c, because of the orientation of the pulse-tube cooler with respect to the direction that the cabin tilts with the elevation angle of the telescope.
With a concentric horizontal optical axis analogous to CONCERTO\cite{CONCERTO2020}, there was insufficient distance between the beam and the wall of the cabin to fit the system. 
This led to the design as shown Fig. \ref{fig:cryostat}a, where the main optical axis is vertical to the cryocooler.

A particular area of challenge was the magnetic shielding.
From the experience of DESHIMA\cite{Karatsu2026DESHIMA2} and AMKID\cite{AMKID_Reyes2026} that use comparable CPW-based NbTiN/Al hybrid KIDs, it was recommended that the terrestrial magnetic field should be attenuated at the IFUs by at least a factor of $2\cdot 10^3$. (This might not necessarily be the case for KIDs using microstrip lines and PPCs, because of the continuous superconducting ground plane.) 
Such a shielding factor was difficult to achieve with a set of concentric tubes made of mu-metal and superconductor.
This led to the design of a compact Nb/mu-metal double-layer magnetic shield as presented in Fig. \ref{fig:mag_shield}.
The COMSOL simulations show the magnetic shielding factor against a field in the direction indicated in Fig. \ref{fig:mag_shield}a, which is the most difficult to shield.
The results in panel b show that the minimum shielding factor is $\sim2.1\cdot 10^3$, meeting the requirements.
The Nb magnetic shield is designed to also serve as a unit that can be dismounted with the IFUs from the cold stage to safely exchange the IFUs as shown in Fig. \ref{fig:cryostat}c.

\section{Summary and Outlook}
\label{sec:summary} 

We presented the conceptual design of TIFUUN: an imaging spectrometer in the 90--360 GHz band with up to 18,000 KIDs, covering the $\diameter$7.5 arcmin field-of-view of ASTE.
Thanks to the superconducting circuit IFU, as well as optimizations in the silicon-lens refractive optics and the dilution-refrigerator cryostat, the system is remarkably compact for an instrument of this scale. 
Indeed, the volume of the cryostat is similar to that of the single-spaxel DESHIMA spectrometer that has just 339 KIDs in 1 spaxel\cite{Karatsu2026DESHIMA2}.
The design of the cryo-mechanical structure is driven by the open-hardware IFU concept, allowing users to easily exchange the IFUs on site.
The ample degrees of freedom in frequency range, spectral resolution, bandwidth, and field-of-view, encourages users to develop astronomical surveys and IFU designs in tandem.
Users are provided with the \texttt{RAIMAD}\cite{RAIMAD} software to design IFUs, and the \texttt{gateau}\cite{gateau,GATEAUgithub} time-domain simulator to jointly optimize the astronomical science, observing strategy, and IFU configuration.
As its first application, TIFUUN targets the SUBLIME experiment on the ASTE 10~m telescope.
In the future, the compact and portable mechanical design makes TIFUUN compatible with emerging telescope facilities such as FYST\cite{FYST} and AtLAST\cite{AtLAST2025,AtLAST2026SPIE}/LST\cite{LSTKohno2020}.
The projected sensitivity of TIFUUN on the AtLAST 50 m can be calculated using the AtLAST sensitivity calculator that is publicly available\cite{atlast_sensitivity_calculator}.
On such facilities, TIFUUN is suited as a first-generation spectral imager, as well as a testbed for fielding IFUs that will continue to improve with the rapid advances in superconducting astrophotonics\cite{AstrophotonicsRoadmap}.

%% Action 3: Add funding information to the acknowledgment
\acknowledgments       

This work was supported by the European Union (ERC Consolidator Grant No.\ 101043486 TIFUUN) and JSPS (KAKENHI Grant Numbers JP22H04939, JP23K20035, and JP24H00004.)
AtLAST has received funding from the European Union's Horizon Europe and Horizon2020 research and innovation programs under grant agreements No.\ 101188037 and No.\ 951815. Views and opinions expressed are however those of the authors only and do not necessarily reflect those of the European Union or European Research Executive Agency. Neither the European Union nor the European Research Executive Agency can be held responsible for them.

\bibliography{tifuun_spie2026} % bibliography data in report.bib
\bibliographystyle{spiebib} % makes bibtex use spiebib.bst

\end{document}